 \documentstyle[prb,aps,multicol,epsfig]{revtex}
\begin{document}
\draft \date{\today} \title
  {Peak effect, vortex-lattice melting-line and
  order--disorder transition
  in conventional and high-$T_c$ superconductors}

\author{Grigorii P.~Mikitik}
  \address{B.~Verkin Institute for Low Temperature Physics \&
   Engineering, National Ukrainian Academy of Sciences,
   Kharkov 61103, Ukraine}
\author{Ernst Helmut Brandt}
\address{Max-Planck-Institut f\"ur Metallforschung,
   D-70506 Stuttgart, Germany}

\maketitle

\begin{abstract}
We investigate the order--disorder transition line from a Bragg
glass to an amorphous vortex glass in the $H-T$ phase diagram of
three-dimensional type-II superconductors taking into account both
pinning-caused and thermal fluctuations of the vortex lattice.
Our approach is based on the Lindemann criterion and on results
of the collective pinning theory and
generalizes previous work of other authors. It is shown that
the shapes of the order--disorder transition line and the vortex
lattice melting curve are determined only by the
Ginzburg number, which characterizes thermal fluctuations, and by
a parameter which describes the strength of the quenched disorder
in the flux-line lattice. In the framework of this unified
approach we obtain
the $H-T$ phase diagrams for both conventional and high-$T_c$
superconductors. Several well-known experimental
results concerning the fishtail effect and the phase diagram of
high-$T_c$ superconductors are naturally explained by assuming
that a peak effect in the critical current density versus $H$
signalizes the order--disorder transition line in superconductors
with point defects.
\end{abstract}
\pacs{PACS numbers: 74.60.Ge, 74.72.Bk}
    \begin{multicols}{2}
    \narrowtext

\section{Introduction}

In type-II superconductors one often
observes\cite{1,2,3,4,5,6,7,8,9,10,11,12,13,14,15,16} a peak effect
(or fishtail effect) in the critical current density
measured as a function of the applied magnetic field $H$ at a
fixed temperature $T$ or as a function of $T$ at fixed $H$. In
conventional low-$T_c$ materials this peak effect mainly occurs at
magnetic fields $H$ near the upper critical field
$H_{c2}(T)$.\cite{1,2,3,4,5}
In the high-$T_c$ ${\rm YBaCuO}$ crystals, the
line of the maximum critical current density, $H_p(T)$, frequently
lies essentially below the irreversibility line in the $H-T$
plane,\cite{6,7,8,9} and in sufficiently perfect crystals it
exhibits {\it nonmonotonic} behavior with
temperature.\cite{10,11,12,13,14,15,16} In these perfect crystals
the line of maximum current density approaches the flux-line
melting line, $H_m(T)$, approximately at the so-called upper
critical point\cite{17} at which the melting line terminates. When
the oxygen deficiency $\delta$ in ${\rm YBa_2Cu_3O_{7-\delta}}$
increases or the crystal becomes less perfect, the end point tends
to the superconducting transition temperature $T_c$ at zero
magnetic field, while $H_p(T)$ becomes a monotonically decreasing
function.\cite{11,16} It is also important to note that at a fixed
oxygen concentration the fishtail effect can disappear in pure
YBaCuO crystals if the distribution of the oxygen vacancies over
the sample becomes uniform.\cite{18}

 At present the origin of the peak effect in low-$T_c$ and high-$T_c$
superconductors is commonly associated with the proliferation of
dislocations in the flux-line lattice.
\cite{1,2,3,4,5,10,11,12,13,14,15,16} At this first order phase
transition\cite{3,19,20,21,22} induced by quenched disorder in the
vortex system, a transformation of a quasiordered Bragg glass
\cite{23} into a disordered amorphous vortex phase occurs.
Although different criteria\cite{3,12,13,14} are used for
determining the exact position of this transition on the
peak-shaped dependence of the critical current density on $H$,
they all lead to qualitatively similar $H-T$ phase diagrams, and,
for definiteness only, we shall imply below that the phase
transition corresponds to the line of the maximum critical current
density, $H_p(T)$.

  A description of this order--disorder phase transition in
high-$T_c$ superconductors was proposed in
Refs.~\onlinecite{24,25,26} using the Lindemann criterion. It was
implied in these papers that the nature of the order--disorder
phase transition is different from the vortex lattice melting
transition, but at the critical point both phase transition lines
merge. Recently it was refined\cite{22} that the upper critical
point does not generally coincide with the point where the
order--disorder line reaches the melting curve, and thus the
melting line has a portion beyond the intersection point. However,
the following should be noted: The results of
Refs.~\onlinecite{24,25,26} for the disorder--induced transition
were obtained in the regime of single vortex pinning\cite{27} when
the Larkin pinning length $L_c$ is less than $L_0=\epsilon a$
where $a=(\Phi_0/H)^{1/2}$ is the spacing between flux lines,
$\Phi_0$ is the flux quantum, and
$\epsilon=\lambda_{ab}/\lambda_c\le 1$ is the anisotropy of the
superconductor ($\lambda_{ab}$, $\lambda_c$ are the London
penetration depths in the $ab$ plane and along the $c$ axis,
respectively). When $T$ increases, the length $L_c$ should exceed
$\epsilon a$ at some temperature which lies on the boundary of the
single vortex pinning regime. At higher temperatures the disorder
was completely neglected in Refs.~\onlinecite{24,25,26}, and only
the melting line of the ideal lattice was derived. Thus, the
behavior of the order--disorder line was not actually investigated
in the high temperature region, and its connection with the
melting line was not established. Besides this, it has remained
unclear why the proliferation of dislocations in the vortex
lattice of high-$T_c$ and low-$T_c$ superconductors leads to
different phase diagrams.

  In the present paper we consider the order--disorder
transition line in the high temperature region and {\it obtain} the
point where the vortex lattice melting and the order--disorder
transition lines merge. It turns out that for a given model of the
vortex pinning, the resulting $H-T$ phase diagram is determined only
by the Ginzburg number $Gi$, which characterizes the thermal
fluctuations, and by a parameter\cite{27} $j_c(0)/j_0(0)$ that
describes the strength of the quenched disorder in the flux-line
lattice at $T=0$ ($j_0$ is the depairing current density and $j_c$
is the critical current density in the single vortex pinning regime;
both are in the $ab$ plane).
For different values of these parameters, phase diagrams are obtained
which are similar to those observed in experiments for low-$T_c$
and high-$T_c$  superconductors. Thus, the results of this
paper provide a unified approach for analyzing $H-T$ phase
diagrams of various superconductors.

 In this paper we consider only magnetic fields exceeding
considerably the lower critical field $H_{c1}$ and thus disregard the
reentrant behavior of the melting transition and do not distinguish
between the magnetic field $H$ and the magnetic induction $B$.
Besides this, we deal only with anisotropic three-dimensional
superconductors, neglecting completely the decoupling of the
superconducting layers. We also assume that ${\bf H}$ is directed
along the $c$ axis. This assumption simplifies the analysis
of the problem, though our final Eqs.~(19) to (25) are
valid for any direction of the magnetic field.

\section{Lindemann criterion}

  We begin with simple estimates which show that the Lindemann
criterion does define the condition for proliferation of
dislocations in the flux-line lattice at the order--disorder
transition. Consider a dislocation network in the lattice. Let a
unit cell of this network have the dimensions $R_d$ and $L_d$ in the
transverse and longitudinal directions to ${\bf H}$, respectively.
A comparison of tilt and shear elastic energies yields that
  $L_d/R_d \sim [c_{44}(1/R_d,1/L_d)/c_{66}]^{1/2}>1$
where $c_{66}$ and $c_{44}(k_{\perp},k_{\parallel})$ are the shear
and nonlocal tilt moduli of the flux-line lattice.\cite{28} The
energy cost for the creation of a dislocation cell is of the order of
\[
 E_d\sim \varepsilon_0 L_d
\]
where $\varepsilon_0=(\Phi_0/4\pi\lambda_{ab})^2$ and
$\lambda_{ab}$ is the London penetration depth for currents in
the $ab$ plane. On the other hand, the elastic energy in the
volume $R_d^2L_d$ is estimated as
\[
E_{el}\sim c_{66}L_du^2(R_d,L_d)
\]
where $u^2(R,L)$ is the correlation function determining the relative
displacement of points in the lattice with  quenched disorder,
 \[
 u(R,L)\equiv \langle[{\bf u}(R,L)-{\bf u}(0,0)]^2\rangle^{1/2}.
 \]
Here ${\bf u}$ is the transverse displacement of a flux line,
$\langle \dots \rangle$ means averaging over both thermal and
quenched disorder, the first coordinate $R$ in ${\bf u}(R,L)$
indicates the position of the flux line in the plane normal to the
applied magnetic field, while the second coordinate $L$ defines the
position of a point on the flux line. Comparing $E_d$ and $E_{el}$
with account of $c_{66}\sim \varepsilon_0/a^2$, one arrives at the
conclusion that a dislocation network can exist in the
lattice if $u(R_d,L_d)\ge a$. In other words, $R_d$ should be greater
than the so called positional correlation length $R_a$ within which
typical relative vortex displacements are of the order of the lattice
spacing $a$. However, this is only a necessary but not sufficient
condition for the existence of dislocations. Displacements generated
by the dislocations facilitate a better adjustment of the vortex
lattice to the quenched disorder.
The relative deformation of the lattice produced by
the dislocations is of the order of $a/R_d$.
Therefore, the smaller $R_d$ is the greater is the gain
$\delta E_{pin}$ in pinning energy $E_{pin}$. Thus, the network
first appears at the smallest possible $R_d$, and we arrive at
the result $R_d\sim R_a$ (and $L_d\sim L_a$) obtained in
Ref.~\onlinecite{22}. The relative magnitude of the
gain, $\delta E_{pin}/E_{pin}$ is determined by the ratio $a/R_a$.
Hence, for this magnitude to become of the
order of unity, $R_a/a$ should decrease to a certain constant $C$,
 \[
 {R_a \over a}=C .
 \]
This criterion for the appearance of dislocations in the flux-line
lattice was obtained in Refs.~\onlinecite{22,29}
(see also Ref.~\onlinecite{1}), and  is
equivalent\cite{30} to the condition \cite{25}:
\begin{equation} 
  u^2(a,0)=c_L^2a^2 ,
\end{equation}
where $c_L$ is the phenomenological Lindemann constant. This
immediately follows from the fact that the ratio of $u(R_a,0)$ to
$u(a,0)$ [i.e., $a/u(a,0)$] is a function of $R_a/a$. Finally,
since $u(a,0)=u(0,L_0)$ at $L_0=\epsilon a$,\cite{27} one more
form of the Lindemann criterion exists:
\begin{equation} 
  u^2(0,L_0)=c_L^2a^2 .
\end{equation}
It is just this form that was used in Refs.~\onlinecite{24,26}.

 Strictly speaking, the values of the constants $C$ and $c_L$
may depend on whether the order--disorder transition occurs in the
single vortex pinning region or in the region of bundle pinning.
However, to understand the essence of the matter, we shall use the
simplest approximation: $c_L$ will be considered as the same
constant for the various regimes of pinning.

\section{The order--disorder line. Simplified approach}

 As well-known,\cite{27,31} thermal fluctuations of the flux-line
lattice lead to a smoothing of the pinning potential and thereby
affect the pinning. This thermal depinning is especially important
for high-$T_c$ superconductors. However, to elucidate possible
types of the order--disorder transition line, in this section we
completely disregard the thermal fluctuations. The influence of the
thermal depinning on the order--disorder line will be analyzed
in Sec.~IV.

\subsection{Region of single vortex pinning}

  As has been mentioned above, the order--disorder line $H_{dis}(T)$
was studied \cite{24,25,26} inside the single-vortex pinning regime
where the Larkin pinning length $L_c$ is less than $L_0=\epsilon a$.
Since $L_0>L_c$, formulas of the random manifold regime\cite{27} for
a single vortex are applicable to calculate the displacement
correlation $u(0,L_0)$,
 \begin{equation}
 u(0,L_0)\approx \xi(L_0/L_c)^{\zeta},
 \end{equation}
where $\xi$ is the coherence length in the $ab$ plane and $\zeta$
is the roughness exponent for a flux line. In Ref.~\onlinecite{26}
the value $\zeta\approx 3/5$ was used, while $\zeta\approx 5/8$ in
Ref.~\onlinecite{24}. Inserting Eq.~(3) into Eq.~(2), we obtain
after simple manipulations:
 \begin{equation} 
 H_{dis}={\Phi_0 c_L^2 \over \xi^2} \left ({c_L L_c\over
 \epsilon \xi}\right )^{\alpha},
 \end{equation}
where $\alpha=2\zeta/(1-\zeta)\approx 3$. Eq.~(4) coincides with
the appropriate formulas of Refs.~\onlinecite{24,25,26}. For
Eq.~(4) to be self-consistent, it is necessary to verify that
$L_0>L_c$ at $H=H_{dis}$ or in other words $H_{dis}<H_{sv}$ where
$H_{sv}=\Phi_0\epsilon^2/L_c^2$ is the boundary of the single
vortex pinning regime.\cite{27} This condition yields:
 \begin{equation} 
 {\epsilon \xi \over L_c}>c_L.
 \end{equation}
If the inequality (5) is not fulfilled, Eq.~(4) is not valid to
describe $H_{dis}$.

 The parameter $\epsilon \xi /L_c$ generally depends on the
temperature $T$. For example, according to simple estimates given
in Appendix A, Eqs.~(A3), (A7), it decreases with $T$ and reaches
zero at $T=T_c$. Moreover, its decrease becomes especially
pronounced if the thermal depinning is taken into account. [In
this case the single vortex collective pinning length $L_c$
increases sharply\cite{27} when $T$ exceeds the characteristic
pinning energy $\tilde T_{dp}^s(T)$, see Appendix A.] Thus, even
if $\epsilon \xi(0)/c_L L_c(0)>1$, the order--disorder line
$H_{dis}(T)$ reaches the boundary of the single vortex pinning
regime, $H_{sv}(T)$, at some temperature $T_1$ defined by the
condition
 \begin{equation} 
  {\epsilon \xi(T_1)\over L_c(T_1)}=c_L,
 \end{equation}
and at $T>T_1$, Eq.~(4) fails.

 It is worth noting that
the parameter $\epsilon \xi/L_c$ appearing in Eqs.~(4) - (6)
and formulas given below characterizes the strength of the
disorder in the flux-line lattice\cite{27} and is expressed
through the critical current density $j_c$ in the single
vortex pinning regime,
\begin{equation} 
  {\epsilon \xi \over L_c}=\left({j_c \over j_0}\right)^{1/2},
\end{equation}
where $j_0$ is the depairing current density.

\subsection{High temperature region}

  At $T>T_1$ the order--disorder transition line lies above
$H_{sv}(T)$. In this region of the $H-T$ plane  small-bundle and
large-bundle regimes of pinning occur.\cite{27} Hence the
transverse collective pinning length $R_c$ exceeds $a$, and to
find $u(a,0)$, the results\cite{27,32,33} may be used, which were
obtained within the framework of the perturbative approach of
Larkin and Ovchinnikov.\cite{32} We have:
 \begin{equation} 
 u^2(a,0)\approx \xi^2(L_0/L_c)^3\left ( {1-h_{sv}
 \over 1-h}\right )^{3/2},
  \end{equation}
where $L_0=\epsilon a$, $h\equiv H/H_{c2}$, $h_{sv}\equiv
H_{sv}/H_{c2}$, the upper critical field $H_{c2}=\Phi_0/2\pi
\xi^2$, and $L_c$ is the {\it single vortex} collective
pinning length.
Note that Eq.~(8) differs from formula (4.17) of Ref.~\onlinecite
{27} by the last factor containing $h$ and $h_{sv}$. This factor
takes into account the possibility that $1-h$ is small; in
Ref.~\onlinecite{27} the correlation function $u^2(R,L)$ is given
without taking account of this possibility. The origin of this
factor is the following. The quantity $u^2$ is proportional to
 $nf_{pin}^2\lambda_{ab}/Hc_{66}^{3/2}$. When $h \to 1$, one has
 $\lambda_{ab}\propto (1-h)^{-{1\over 2}}$,
 $c_{66} \propto (1-h)^2$ (see Ref.~\onlinecite{28,34}), while
 $f_{pin}^2\propto \varepsilon_0^2 \propto (1-h)^2$
(see Ref.~\onlinecite{32} and also Appendix A).
The combination of these factors gives Eq.~(8), in which the
additional constant factor $(1-h_{sv})^{3/2}$ has been introduced
to provide a smooth crossover of this expression to the appropriate
formula for the single vortex pinning regime at $h=h_{sv}$.

 Inserting formula (8) into Eq.~(1), we obtain an equation for
 $h_{dis}=H_{dis}/H_{c2}$,
 \begin{equation} 
 h_{dis}(1-h_{dis})^3=2\pi c_L^2 \left (\epsilon\  \xi
 \over c_L L_c\right )^6 (1-h_{sv})^3,
  \end{equation}
where
 $h_{sv}=H_{sv}/H_{c2}=2\pi(\epsilon\xi/L_c)^2$ and the right hand
side depends only on the temperature. Note that at $T=T_1$, when
$\epsilon \xi/L_c = c_L$, one has $h_{dis}(T_1)=2\pi
c_L^2=h_{sv}(T_1)$, in agreement with Eq.~(4). A simple analysis
shows that $2\pi c_L^2$ should be greater than $0.25$ but less
than $1$ (i.e., $0.2\le c_L\le 0.4$) for Eq.~(9) to have a solution
at $T\ge T_1$. If $c_L<0.2$, the order--disorder line terminates
at $T=T_1$, which is impossible.\cite{35} On the other hand, if
$c_L>0.4$, one finds from Eq.~(9) that $h_{dis}>1-h_{sv}$, i.e.,
the root of the equation lies in the upper region of single vortex
pinning\cite{36} where Eq.~(9) is not valid. For this reason,
in the following we assume the conditions $0.2\le c_L\le 0.4$ to be
fulfilled and, for definiteness, take $c_L=0.25$ in the subsequent
calculations.

\subsection{Types of phase diagrams}

 We begin the analysis of phase diagrams with the case
\[
 D>c_L
\]
where $D\equiv \epsilon \xi(0)/L_c(0)$ is the value of the
parameter $\epsilon \xi/L_c$  at $T=0$. Figure 1 shows the line
$H_{dis}(T)$ calculated by solving Eq.~(9) (for $H_{dis}>H_{sv}$)
and Eq.~(4) (for $H_{dis}<H_{sv}$). In the construction of this
figure, as well as in all examples below, we use
 $H_{c2}(T)=H_{c2}(0)[1-(T/T_c)^2]$ and $\alpha=3$
(i.e., $\zeta=3/5$). Besides this, taking into account the
formula (A13) of Appendix A, we employ
the following simple approximation for the parameter
 $\epsilon \xi/L_c$:
 \begin{equation} 
 {\epsilon \xi(T)\over L_c(T)}\equiv Dg_0(t)
 \end{equation}
with
 \begin{equation} 
 g_0(t)=(1-t^2)^{1/2},
 \end{equation}
where $t\equiv T/T_c$.
The increase of $H_{dis}$ and its subsequent maximum are seen
in the vicinity of $T_1$. As to $h_{dis}$, this normalized quantity
increases monotonically above $T_1$.
When $1-h_{dis}\ll 1$, an approximate solution of Eq.~(9) is:
 \begin{equation} 
 {H_{dis}(T)\over H_{c2}(T)}\approx 1-
 \left ({2\pi \over c_L^4}\right )^{\!1/3} \left (\epsilon \xi
 \over L_c\right )^{\!2}.
  \end{equation}
This formula shows that and how $H_{dis}(T)$ approaches
$H_{c2}(T)$, which is also seen in Fig.~1. Interestingly, according
to this formula, the order--disorder transition occurs
{\it outside} the upper region of single vortex pinning\cite{36},
but its position correlates with the boundary of this region:
$[1-h_{dis}(t)]/[1-h_{sv}(t)]=(2\pi c_L^2)^{-2/3}>1$.
It should be also noted that
in the case under study (i.e.\ when the thermal fluctuations are
negligible) the mean-field $H_{c2}(T)$ practically
coincides\cite{37} with the melting line $H_m(T)$. Thus, we obtain
for $D>c_L$ that in the high temperature region a peak effect
occurs near the melting line, while with decreasing $T$ the
position of the peak in $j_c(H)$ shifts downwards from this line.
This situation is reminiscent of that of perfect high-$T_c$
superconductors.\cite{10,11,12,13,14,15,16}

  In this context it is also useful to note the following:
The density of the dislocations in the vortex liquid is
essentially higher than in the disordered vortex solid phase near
the order--disorder transition.\cite{22} However, if in the $H-T$
plane the order--disorder transition occurs sufficiently below the
melting line of the clean superconductor, then in the disordered
solid phase, at the field corresponding to the melting transition,
the density of the dislocations generated by the quenched disorder
may become of the order of the density characteristic for the
liquid phase. In this case the melting transition {\it
disappears}. In other words, the melting line $H_m(T)$ terminates
when $H_{dis}(t)$ deviates from it appreciably.

 If the strength of the disorder is sufficiently small,
\[
  D\equiv {\epsilon \xi(0) \over L_c(0)}<c_L ,
\]
the order--disorder line lies entirely outside the region of
single vortex pinning, Fig.~2, and is described by Eq.~(9) at any
$T<T_c$, while Eq.~(4) is not valid at all. In the special
situation when $D$ is markedly less than $c_L$, a peak effect
occurs near $H_{c2}(T)$ and its position in the $H-T$ plane is
approximately given by Eq.~(12). In this case the resulting phase
diagram looks like that of low-$T_c$
superconductors.\cite{1,2,3,4,5} The transition from one type of
phase diagram to the other occurs when $D=c_L$.

 It has been assumed in this section that the parameter
$\epsilon \xi/L_c$ decreases with increasing $T$. However, the
$\delta T_c$ pinning (due to spatial variations of $T_c$)
leads to an increasing function $g_0(t)$, see
Eq.~(A14) in Appendix A. In this case, if $D>c_L$, the formula (4)
remains valid up to $T_c$. But if $D<c_L$, a temperature $T_0$
exists, determined by the condition
 \[
  {\epsilon \xi(T_0)\over L_c(T_0)}=c_L \,,
 \]
and at $T<T_0$ equation (9) should
be used, while at $T>T_0$ formula (4) holds. In other words, we
have a situation which is opposite to that described above.
In Fig.~3 the
order--disorder line is shown for the case when $g_0(t)$ is given
by Eq.~(A14). Note that in this case, according to formula (4),
one has $H_{dis}\propto (1-t^2)^{3/2}$ in the high temperature
region of the phase diagram. This result qualitatively agrees
with the measurements\cite{6,7,8,9,11,16} on
${\rm YBa_2Cu_3O_{7-\delta}}$ crystals when $\delta$ is not small,
or when the crystals are not too perfect.

\section{The order--disorder line with account of thermal
fluctuations}

  It is well-known that thermal fluctuations of the flux-line
lattice play an important role in high-$T_c$ superconductors. In
particular, for this reason the flux-line lattice melts
essentially below the mean-field $H_{c2}$ line. In this section
we study the influence of thermal fluctuations on the
order--disorder line.

  Thermal fluctuations lead to a smoothing of the pinning
potential and thus increase the Larkin length $L_c$. The length
$L_c$ specifies the boundary of the single vortex pinning region
and it enters the key parameter of the collective pinning theory,
$\epsilon \xi /L_c$. Here we reserve the notation $L_c$ for the
true length renormalized by the fluctuations, while the Larkin
length defined without account of the fluctuations will be denoted
below as $L_c^0$. Note that just $L_c^0$ has been used in
Sec.~III, and just this quantity is described by Eq.~(10). Apart
from increasing $L_c$, the thermal fluctuations of the flux-line
lattice also modify the correlation function (3) as
follows\cite{27}:
 \begin{equation} 
 u^2(0,L_0)\approx r_p^2(L_0/L_c)^{2\zeta},
 \end{equation}
where $r_p^2=\xi^2+u_T^2$, and $u_T$ is the magnitude of these
fluctuations, which depends on the temperature and on the magnetic
field. It is implied in Eq.~(13) that
$L_0=\epsilon a>L_c$. As to the correlation function (8), the
collective pinning theory\cite{27} gives
 \begin{equation} 
 u^2(a,0)\approx \xi^2(L_0/L_c^0)^3\left ( {1-h_{sv}
 \over 1-h}\right )^{\!3/2} \! \left({\xi \over r_p}\right)^{\!4}
  \end{equation}
for $L_0<L_c$, i.e.\ at $h>h_{sv}=2\pi (\epsilon \xi /L_c)^2$.

  Since $u(a,0)=u(0,L_0)$, the correlation functions (13) and (14)
must coincide at $L_0=L_c$ (or equivalently at $h=h_{sv}$).
This condition yields
 \begin{equation} 
  L_c(t)=L_c^0(t)\,{r_p^2(t,H_{sv}(t)) \over \xi^2(t)}.
 \end{equation}
In fact, formula (15) is an {\it equation for} $L_c$ since we have
the relationship $H_{sv}=\Phi_0\epsilon^2/L_c^2$. This equation
enables us to find $L_c(t)$, and thus $H_{sv}(t)$,
self-consistently.

 To proceed, we have to estimate the magnitude of
thermal displacements of the lattice, $u_T$, relative to its
equilibrium position. In the case of the {\it ideal} vortex
lattice, $u_T$ was calculated in many papers; see, e.g.,
Refs.~\onlinecite{38,39,40}. This magnitude, as well as the
correlation functions (8) and (14), depends on the elastic moduli
of the lattice. However, in deriving Eqs.~(8), (14) the
contribution associated with the compression modulus $c_{11}$ was
neglected. Hence, it is consistent to use the same approximation
in the calculation of $u_T^2$. This simplifies the appropriate
formula\cite{40} for $u_T^2$, and we obtain
\begin{equation} 
  u_T^2\approx \xi^2 t
  \left ({Gi\over 1-t^2} \! \right)^{\!1/2}h^{-1/2}f(h),
\end{equation}
where $h=H/H_{c2}(t)$, $t=T/T_c$, $H_{c2}(t)=H_{c2}(0)(1-t^2)$,
$Gi\,$ is the Ginzburg number,
\[
  Gi={1\over 2}\left ({T_c \over H_c^2(0)\epsilon \xi^3(0)}
  \right)^{\!2},
\]
which characterizes the strength of the thermal fluctuations, and
$H_c$ is the thermodynamic magnetic field of the superconductor. The
function $f(h)$ has the form:
  \begin{equation} 
  f(h)= {2\beta_A \over 1-h} {[1+(1+\tilde c)^2]^{1/2}-1
  \over \tilde c(1+\tilde c)},
  \end{equation}
with $\tilde c=0.5[\beta_A(1-h)]^{1/2}$, and $\beta_A=1.16$.

 The quenched disorder changes $u_T^2$. However, when the
transverse collective pinning length $R_c$ is considerably greater
than $a$, the above result for the ideal lattice is a good
approximation. This is due to the fact that the main contribution
to $u_T$ results from the thermal fluctuations with short
wavelengths ($k_{\perp}\sim 1/a$), while the quenched disorder
essentially distorts the lattice only on the scale $R_c$. Thus, we
may use Eq.~(16) in the case of the {\it nonideal} lattice if
bundle pinning occurs. But in the single vortex pinning regime the
influence of the disorder is essential, and one has\cite{27}
$u_T^2\propto L_c$. To account for this result, we introduce an
additional factor $L_c/\epsilon a$ in the formula (16) and thus
obtain
\begin{equation} 
  u_T^2\approx \xi^2 t
  \left ({Gi\over 1-t^2} \right)^{\!1/2} \!h_{sv}^{-1/2}f(h)
\end{equation}
in the single vortex pinning region at $h<h_{sv}$.

 Inserting expression (16) into formula (15) and using
definition (10) for $\epsilon \xi/L_c^0$, we obtain the
following equation for $h_{sv}=H_{sv}/H_{c2}$:
\begin{equation} 
  h_{sv}^{1/2}(t)=(2\pi )^{1/2}Dg_0(t) -
  t \left ({Gi\over 1-t^2} \right)^{\!1/2} \!\! f(h_{sv}(t)).
\end{equation}
With increasing $t$
the function $h_{sv}(t)$ reaches zero at some $t_{dp}^s<1$, and
Eq.~(19) is valid at $t\le t_{dp}^s$. In the region $t>t_{dp}^s$ the
length $L_c$ is infinite in size, and the single vortex
pinning regime is absent, i.e., $h_{sv}(t)=0$ at $t>t_{dp}^s$.
The value of $t_{dp}^s$ is found by equating $h_{sv}$ to zero
in formula (19):
 \begin{equation} 
  t_{dp}^s={(2\pi)^{1/2}D\over
  Gi^{1/2}f(0)}g_0(t_{dp}^s)[1-(t_{dp}^s)^2]^{1/2}.
 \end{equation}
It may be verified that the right hand side of Eq.~(20) coincides
(up to a numerical factor of the order of unity) with the
dimensionless characteristic pinning energy $\tilde T_{dp}^s/T_c$,
see Appendix A. Thus, in agreement with physical
considerations,\cite{27} we obtain that the essential
renormalization of $L_c$ occurs at such temperatures $T$ that
$T\sim \tilde T_{dp}^s(T)$.

  It should be noted that our result for $L_c$ ($L_c \to \infty$
at $t \to t_{dp}^s$) differs in some respects from that presented
in Ref.~\onlinecite{27} where $L_c$ increases exponentially at
$t\sim t_{dp}^s$. However, in the framework of our approximation,
$H_{c1}=0$, we may consider $L_c$ as infinite if it becomes
of the order of $\lambda$. Hence, the difference between the
results is, in fact, small. But our approach provides the
continuity of the correlation functions (13), (14) at $h=h_{sv}$.

  Equation (19) specifies the single vortex pinning (SVP) region
existing at relatively low magnetic fields [$0<h<h_{sv}(t)$].
However, formulas (14), (16) enable one to find also the upper
region in which the vortex system returns to this type of pinning
again. The appropriate equation results from the condition
$u(a,0)=r_p$, and has the following form:
\begin{eqnarray} 
 (1-h) \left[h^{1/2} +
 t \left ({Gi\over 1-t^2} \right)^{\!1/2} \!\!f(h)\right]^2
 = \nonumber \\
  2\pi  \left (D g_0(t) \right )^{\!2} [1-h_{sv}(t)],
\end{eqnarray}
As might be expected, at $h=h_{sv}$ this equation goes over into
Eq.~(19). However, in a certain temperature interval it has two
additional real roots which form the boundary of the upper SVP
region, $h_{sv}^{up}$ (see Fig.~5 below). Here we do not consider
this issue in detail, but qualitatively describe the effect of
thermal fluctuations on the shape of the upper SVP region
specified above without their account.\cite{36} Although the
softening of the vortex lattice near $H_{c2}$ is favorable for
single vortex pinning, this softness also leads to an increase of
the thermal fluctuations $u_T$, which reduces the strength of
pinning. As a result, the upper region does not touch $H_{c2}(t)$
except for the point $t=0$ (at $t>0$ and $H=H_{c2}$ we have
$u_T=\infty $). Besides this, since $u_T$ increases with $t$, the
upper region does not extend to $T_c$. Of course, one should
realize that the part of the boundary of this SVP region lying
above the vortex lattice melting curve has only formal meaning,
since it does not account for the vanishing shear modulus at the
melting. On the other hand, it is quite possible that the sharp
melting transition disappears when the melting line enters the
upper SVP region (or even before it; see Sec.~V).

  Inserting expressions (13) and (18) in formula (2), we arrive
at an equation for $h_{dis}=H_{dis}/H_{c2}$ which generalizes
formula (4):
 \begin{eqnarray} 
 h_{dis} \left[1 +
  t \left ({Gi\over 1-t^2} \right)^{\!1/2} \!\!{f(h_{dis})\over
  (h_{sv}(t))^{1/2}}\right]^{1/(1-\zeta)}= \nonumber \\
 2\pi c_L^2\left({2\pi c_L^2\over
 h_{sv}(t)}\right)^{\!\alpha/2} .
  \end{eqnarray}
This equation is valid in the single vortex pinning regime when
$h_{dis}\le h_{sv}$. For example, if $g_0(t)$ is a decreasing
function of $t$, the order--disorder line lies in the single
vortex pinning region and is described by Eq.~(22) at $t<t_1$. In
the case $D>c_L$ the temperature $t_1$ is found from the equation:
 \begin{equation} 
  h_{sv}(t_1)=2\pi c_L^2 \left[1 +
  t \left ({Gi\over 1-t^2} \right)^{\!1/2} \!\!{f(h_{sv}(t))\over
  (h_{sv}(t))^{1/2}}\right]_{t=t_1}^{-1} \!\!,
\end{equation}
which generalizes condition (6). At $D<c_L$ the line is entirely
outside this region, and thus one has $t_1=0$. If $g_0(t)$ is an
increasing function of temperature, more complicated situations
can occur.

 Inserting formulas (14) and (16) into relation (1), we obtain
the equation for $h_{dis}(t)$:
  \begin{eqnarray}  
 h_{dis}(1-h_{dis})^3 \left[1 +
  t \left ({Gi\over 1-t^2} \right)^{\!1/2} \!\!{f(h_{dis})\over
  (h_{dis})^{1/2}}\right]^4 = \nonumber \\
 2\pi c_L^2 \left ({D g_0(t)
 \over c_L }\right )^{\!6} [1-h_{sv}(t)]^3,
  \end{eqnarray}
which generalizes Eq.~(9). This equation is valid in the bundle
pinning region.

  Let us now present formulas for the melting line which is
determined by the Lindemann criterion, $u_T^2=c_L^2a^2$, {\it
different} from Eqs. (1), (2). This well-known empirical criterion
based on the magnitude of the thermal fluctuations,
was justified in Ref.~\onlinecite{22} for the case of the ideal
vortex lattice. According to this paper, different physical
mechanisms lead to the proliferation of dislocations at the vortex
lattice melting and at the order--disorder transition. While the
disorder--induced transition is driven by an adjustment of the
flux-line lattice to the disorder, the thermal melting is governed
by the entropy gain associated with the creation of dislocations.
Hence, the Lindemann constant $c_L$ for the melting may, in
principle, differ from that used in Eqs.~(22)-(24). However, for
the sake of simplicity we take these constants $c_L$ as equal in
the following analysis. Thus, if the melting line $h_m(t) = H_m
/H_{c2}$  does not intersect $h_{sv}(t)$, it is described by the
equation:
\begin{equation} 
  t \left ({Gi\over 1-t^2} \right)^{\!1/2} \!h_m^{1/2}
  \,f(h_m)=2\pi c_L^2.
\end{equation}
But if the melting line lies below $h_{sv}(t)$, an additional
factor $(h_m/h_{sv})^{1/2}$ should be inserted on the left hand
side of Eq.~(25), cf.\ Eqs.~(16) and (18). Finally, we
note that in expression (25), as well as in Eqs.~(16),
(18)-(24), the factor $1-t^2$ represents the temperature
dependence of the upper critical field,
$1-t^2=H_{c2}(t)/H_{c2}(0)$. Thus, if another form of this
dependence is implied, the appropriate modification of all these
equations is straightforward.

  Summing up, we may state the following: For given $g_0(t)$,
equations (19)-(25) enable us to calculate the order--disorder
line $h_{dis}(t)$, the melting line $h_m(t)$ and the boundaries of
the single vortex pinning regime $h_{sv}(t)$, $h_{sv}^{up}(t)$
with account of the thermal fluctuations. The function $g_0(t)$ in
Eq.~(10) is determined by the pinning mechanism and by the
temperature dependences of $\xi$ and $\lambda$, see Appendix A.

\section{Analysis of phase diagrams}

 It is important to emphasize that for given $g_0(t)$
Eqs.~(19)-(25) depend only on {\it two} parameters: the strength
of quenched disorder, $D = \epsilon \xi(0)/L_c(0)$, and the
strength of the thermal fluctuations, $Gi$. In this sense, the
equations and figures of Sec.~III correspond to the limiting case
$Gi \to 0$. We consider now new features of the phase diagram
which appear in the real situation of finite $Gi$.

\subsection{Numerical results}

  An example of the phase diagram in the case $D>c_L$ is shown
in Fig.~4. Note that the order--disorder line terminates at some
temperature $t_e<1$. This termination is associated with an
increase of the thermal fluctuations and thus with an enhanced
smoothing of the pinning potential when the order--disorder line
approaches $H_{c2}$. Interestingly, the end point lies near the
so-called depinning line\cite{27,41} where $u_T^2=\xi^2$. Another
new feature of the phase diagram is the intersection of the
melting and the order--disorder lines. Thus, we obtain the point
where both lines merge. It is seen from Fig.~4 that at this point
the position of the critical current peak begins to shift sharply
downward from the melting curve, in agreement with the
experimental data.\cite{10,11,12,13,14,15,16} Hence, as was
mentioned in Sec.~III C, the upper critical point for the melting
is likely to occur somewhere nearby. The portion of the
order--disorder line lying above the melting curve has no physical
meaning since the density of dislocations in the liquid phase is
already higher than in the disordered solid phase.\cite{22} Note
also that the rise of $H_{dis}(t)$ at $T>T_1$ becomes considerably
steeper than in the case $Gi=0$.

  The intersection of the order--disorder line
with the melting curve occurs also for $D<c_L$. In this case,
when $Gi$ increases, the decrease of $H_{dis}$ with $T$ diminishes,
and eventually $H_{dis}$ becomes an increasing function of $t$,
see Fig.~5. Therefore, if $Gi\sim 10^{-2}$ (this value is
typical for ${\rm YBaCuO}$ crystals), the temperature
behavior of the order--disorder line is similar to that obtained
for $D>c_L$, and the phase diagrams in both these cases are of
the same type.

  In Fig.~5 we also show the upper region of single vortex
pinning. It is important to note that the intersection point of
the order--disorder line and the melting line occurs clearly {\it
before} the melting curve enters this region, and so at the
intersection point the transverse collective pinning length $R_c$
and the positional correlation length $R_a$
(the dislocation spacing) both considerably
exceed the flux-line spacing $a$. This fact justifies the assumed
weak influence of quenched disorder on the vortex lattice melting
and the application of Eq.~(24) up to the intersection point.

  If the parameter $D$ increases, e.g., as a result of
irradiation or of reduction of the oxygen content in ${\rm
YBa_2Cu_3O_{7-\delta}}$ crystals (see Appendix A), then
the order--disorder line shifts down, while the temperature
of the intersection point (and thus the temperature of the upper
critical point for the melting line) increases, Fig.~6. These
results are in agreement with the experimental
findings\cite{10,11,15,16} for ${\rm YBa_2Cu_3O_{7-\delta}}$
crystals. However, although the obtained results qualitatively
describe the phase diagram of these crystals and its evolution
with varying $D$, it should be emphasized that the intersection
point does not reach $T_c$ at reasonable values of the disorder
parameter $D$.

  Consider now the case when $g_0(t)$ increases with $t$. This
situation occurs in the model of $\delta T_c$ pinning, see
Appendix A. The appropriate phase diagrams are presented in
Figs.~7-9. It is important to note the following: At sufficiently
strong disorder, the line $H_{dis}(t)$ monotonically decreases
with temperature and practically reaches $T_c$. When the strength
of the disorder decreases, the order--disorder line exhibits
nonmonotonic behavior with $t$, and the temperature of the
intersection point, $t_i$, goes down. Thus, in contrast to the
case of a decreasing function $g_0(t)$, the model of $\delta T_c$
pinning correctly reproduces all the features of the experimental
data for ${\rm YBa_2Cu_3O_{7-\delta}}$
crystals.\cite{6,7,8,9,10,11,12,13,14,15,16} In particular, the
results presented in Fig.~8 closely resemble the development of
the order--disorder line with variations of the oxygen deficiency
$\delta$, see, e.g., Fig.~9 in Ref.~\onlinecite{11}.

 The results of this section may also provide an explanation of the
findings obtained in Ref.~\onlinecite{18}. It was shown in this
paper that the flux-line pinning in pure ${\rm
YBa_2Cu_3O_{7-\delta}}$ crystals is mainly due to a nonuniform
distribution of oxygen vacancies over the sample. Changing the
conditions of annealing of the sample, Erb et al. changed this
distribution at a fixed $\delta$. When the distribution became
more homogeneous, the fishtail effect disappeared. Since the
spatial fluctuations in the density of oxygen vacancies strengthen
the $\delta T_c$ pinning, we get an additional confirmation of the
hypothesis that this type of pinning plays the main role in not
too perfect ${\rm YBaCuO}$ crystals, and so the fishtail effect
exists up to high temperatures, as shown in Figs.~7, 8 for large
$D$. When, after annealing, the crystal becomes more perfect, and
the above-mentioned spatial fluctuations are reduced, the value of
the parameter $D$ appears to decrease considerably. (Moreover, it
is conceivable that some other type of pinning begins to
dominate.) Thus, we arrive at the situation when a fishtail effect
is absent, at least in the region of not too high magnetic fields,
as it follows from the data shown in Figs.~6 and 8 for small $D$.

 Finally, we briefly describe the case of small
Ginzburg numbers which occurs for conventional superconductors.
In this case the temperatures $t_{dp}^s$ and $t_i$ are
practically equal to unity, the melting line almost coincides
with $H_{c2}(t)$, and the phase diagrams tend to those shown in
Figs.~1-3. Thus, the results of the simplified approach of
Sec.~III can be well applied to such superconductors.

\subsection{Approximate formulas. Discussion}

  We now present analytical results which give some insight
into the origin of the above-mentioned features of the phase
diagrams. Let us start with the melting line. If
$[Gi/(1-t^2)]^{1/2}\ll 2\pi c_L^2$, the normalized field $h_m =
H_m/H_{c2}$ is close to unity, and we obtain from Eq.~(25)
\begin{equation} 
 1-h_m\approx \left({f_1(1)\over 2\pi c_L^2}\right)^{\!2/3} \!
 t^{2/3}  \left({Gi \over 1-t^2}\right)^{\!1/3}  \!,
\end{equation}
where
\[
f_1(h)\equiv (1-h)^{3/2}f(h)
\]
is defined through the function $f(h)$ given by Eq.~(17), with
$f_1(1)\approx 1.78$. Equation (26) agrees with the strict
result\cite{37} derived for clean superconductors without using
the Lindemann criterion. Formula (26) expressed in usual units
($H_m=h_mH_{c2}$) means that
\[
 H_{c2}(t)-H_m(t) \propto Gi^{1/3}(1-t^2)^{2/3}.
\]
The right hand side of this expression is the width of the
fluctuation region in not too small magnetic fields,\cite{42,43}
$h\gg Gi$. In the opposite limiting case when
$f_1(0)[Gi/(1-t^2)]^{1/2}\gg 2\pi c_L^2$ [but the temperature is
outside the critical region for zero magnetic field,
$Gi<(1-t^2)$], the field $h_m$ is small, and Eq.~(25) yields:
\begin{equation} 
 h_m=\left({2\pi c_L^2 \over f_1(0)t}\right)^{\!2} {1-t^2
 \over Gi},
\end{equation}
where $f_1(0)\approx 2.34$. In fact, this is the well-known
result\cite{27,38,39,40} for clean superconductors,
\[
H_m\propto (1-t)^2.
\]
Note that for Eq.~(27) to hold in a sufficiently wide temperature
region, the Ginzburg number should not be too small. Finally, we
point out an interesting feature of Eq.~(25) that follows from our
numerical analysis. The formula
\begin{equation} 
 H_m= A (1-t^2)^{\gamma} ,
\end{equation}
with some constants $A$ and $\gamma$ turns out to give a very good
fit to $H_m(t)$ determined by Eq.~(25) in the wide temperature
interval $0.5 < t <0.98$, see Fig.~10. The values of the
parameters $A$ and $\gamma$ depend on $Gi$, and the exponent
$\gamma$ increases with increasing $Gi$. In particular, we find
$1.24<\gamma <1.59$ when $0.001<Gi<0.01$. In this connection it is
worth noting that formulas of the type of Eq.~(28) are widely used
to approximate experimental data, and frequently an exponent
$\gamma \approx 4/3$ is found, which is characteristic for the
fluctuation region of a 3D $XY$-type phase transition. We
emphasize here that a good fit
by such formulas does not necessarily mean the existence of a
large fluctuation region in zero magnetic field, but may result
from the specific form of the expression [Eq.~(25)] describing the
melting line.

  In the case of $\delta T_c$ pinning the melting line may lie inside
the region of single vortex pinning, i.e.\ the inequality
$h_{sv}>h_m$ may hold. This occurs if the parameter $\nu$
(see Ref.~\onlinecite{27}, p.~1218),
\begin{equation} 
 \nu \equiv {(2\pi)^{3/2} D^3 \over Gi^{1/2}},
\end{equation}
is sufficiently large, $\nu^{1/3}\gg 1$. Although
in this case the shape of the melting line requires a special
investigation and is not discussed in detail here,
it should be realized that the sharp melting transition
may disappear in this region.

  Let us now analyze Eq.~(24) which describes the line $h_{dis}(t)$
in the region $h>h_{sv}$. To understand the behavior of this
line, we note the following: The function $f_1(h)=(1-h)^{3/2}f(h)$
decreases monotonically with increasing $h$; its variation in the
interval $0<h<1$ is not large, $f_1(1)/f_1(0)\approx 0.76$, and so
if one considers $f_1(h)$ as constant, $f_1(h)=f_1(0)\equiv
f_1\approx 2.34$, this leads to a sufficiently accurate
approximation in solving Eq.~(24). In this approximation Eq.~(24)
becomes a quadratic equation in the variable
$h_{dis}^{1/2}(1-h_{dis})^{3/2}$, and its solution has the form:
\begin{eqnarray}  
 h_{dis}^{1/2}(1-h_{dis})^{3/2} =  \nonumber \\
 F_D-F_T + [ (F_D-F_T)^2-(F_T)^2 ]^{1/2}
\end{eqnarray}
where $F_D$ and $F_T$ are the following functions of temperature:
\begin{eqnarray}
 F_D(t) ={(2\pi)^{1/2}[Dg_0(t)]^3 \over 2c_L^2}\,
 [1-h_{sv}(t)]^{3/2}\,,  \nonumber \\ \nonumber
 F_T(t) =f_1\, t \left({Gi \over 1-t^2} \right)^{1/2} .
\end{eqnarray}
At $t=0$ we have $F_T(0)=0$, and equation (30) goes over into
Eq.(9). Thus, the effect of thermal fluctuations is reduced to a
{\it renormalization} of the right hand side of Eq.~(9) [including
the change of the function $h_{sv}(t)$ according to Eq.~(19)].
Note that the magnitude of this renormalization (i.e., the ratio
$F_T/F_D$) is mainly determined by the parameter $\nu^{-1}$, see
Eq.~(29). Interestingly, in contrast to the case of the $\delta
T_c$ pinning where $\nu$ appears as a result of the specific
temperature dependence of $g_0$, Eq.~(A14), we now conclude that
the parameter $\nu$ characterizes the relative strength of pinning
of any type.

  It should be emphasized that the right hand side of Eq.~(30) is
real only if $F_D\ge 2F_T$. Thus, we find that the order--disorder
line terminates at a temperature $t_e$ which satisfies the
equation
\begin{equation} 
 F_D(t_e)=2F_T(t_e).
\end{equation}
Taking into account Eqs.~(16) and (31), it is easy to determine
the ratio $u_T^2/\xi^2$ at the end point of the order--disorder
line, $t=t_e$, $h=h_{dis}(t_e)$:
\[
 { u_T^2 \over \xi^2} =F_T(t_e)h^{-1/2}(1\!-\!h)^{-3/2} =
 { F_T(t_e) \over F_T(t_e) } =1 \,.
\]
Hence, this end point lies on the so-called depinning line\cite{27}
defined by the condition $u_T^2=\xi^2$, see Figs.~4, 5.
It can be also verified that $t_e$ is always less than the
temperature $t_{dp}^s$ determined by Eq.~(20).

  As has already been mentioned above, only the intersection
point of the melting and the order--disorder lines has a physical
meaning rather than the end point of $H_{dis}(t)$. The temperature
of this intersection point, $t_i$, can be simply estimated using
formulas (26) and (30). Then we arrive at the following equation
for $t_i$:
\[
 4\pi c_L^2 F_D(t_i)=(1+2\pi c_L^2)^2 F_T(t_i) ,
\]
or explicitly,
\begin{equation} 
 t_i=\nu \,{[g_0(t_i)]^3 (1-t_i^2)^{1/2}[1-h_{sv}(t_i)]^{3/2}
 \over f_1(1+2\pi c_L^2)^2}\,.
\end{equation}
This equation suggests that the temperature of the intersection
point, $t_i$, depends on the parameters $Gi$ and $D$ mainly
through their combination $\nu$, Eq.~(29). The data of Fig.~11
support this
hypothesis, viz., at given $g_0(t)$ the temperatures $t_i$
calculated for various $Gi$ fall on the same curve. Of course,
this prediction, as well as any {\it quantitative} conclusion
based on Eqs.~(19)-(25), requires an experimental verification
since a number of simplifying assumptions were made above. In
particular, $c_L$ was assumed to be the same constant for the
melting and for the order--disorder transition in all the pinning
regimes. On the other hand, according to Eq.~(32), the dependence
of $t_i$ on $c_L$ is relatively weak.

  Since $t_i$ specifies the width of the temperature region where
the order--disorder transition exists, the data of Fig.~11 mean
that this width is characterized by the parameter $\nu$. These
data also shed light on the different behavior of $H_{dis}(t)$ in
Figs.~6 and 8 at large values of $D$. To elucidate the results of
Fig.~11 and the conclusions made on their basis, let us analyze
Eq.~(32) in two limiting cases. If $\nu \ll 1$, one obtains the
following estimate from this equation:
\[
t_i\approx 0.42(1+2\pi c_L^2)^{-2}\nu \,.
\]
In other words, we have a situation qualitatively similar to that
shown in Fig.~6 or in Fig.~8 for the smallest $D$. In this case
the order--disorder transition occurs only at very low
temperatures $t<t_i$, while the portion of the melting line at
$t>t_i$ has a large extension. Note that, according to the
estimates of $Gi$ and $D$ presented in Ref.~\onlinecite{2}, such a
situation must take place in pure crystals of 2H-NbSe$_2$, which
were investigated in numerous papers. This conclusion does not
contradict the observation of a peak effect in these
crystals,\cite{2} since a peak effect can signalize not only the
order--disorder transition, as it has been implied so far, but
also the vortex lattice melting.\cite{44,45,46,47,48} Of course,
the features of the peak effect may differ in these two cases.

In the opposite limit, $\nu \gg 1$, the order--disorder transition
reaches $T_c$, while the region of ``pure'' melting ($t_i<t<1$) is
contracted. This limit just corresponds to conventional
superconductors for which $Gi \ll 1$, see Figs.~1-3. However, in
the process of approaching this limit, the evolution of the phase
diagram is different for the different models of pinning, i.e., in
the cases of Eqs.~(A13) and (A14). That is why Figs.~6 and 8
differ from each other at large values of $D$. If Eq.~(A13) is
valid, one finds from Eq.~(32) that
\[
 1-t_i^2\approx 1.5(1+2\pi c_L^2)\nu^{-1/2}.
\]
Thus, a large $\nu$ is required for the temperature $t_i$ to reach
unity, see Fig.~6. In the case of Eq.~(A14) the right hand side of
Eq.~(32) already exceeds unity at a finite value $\nu$ ($\nu \sim
1$), and at such $\nu$ the line $H_{dis}(t)$ practically reaches
the point $T=T_c$, $H=0$, see Fig.~8 for the largest $D$.

\subsection{Upper critical point.}  

  Let us now discuss the upper critical point on the vortex
lattice melting curve.\cite{17} In principle, two scenarios are
possible. In the first one\cite{22} the upper critical point does
not coincide with the intersection point of the melting and the
order--disorder lines and is located at a higher magnetic field
than this intersection point. In fact, in this case the intersection
of the {\it two different} phase transition lines occurs, and one of
them (the order--disorder line) terminates at the intersection
while the other one continues for some distance which seems to
depend on the detailed position of the lines in the $H-T$ plane.
This scenario is
based on the difference in the density of dislocations, $\rho$,
generated at the melting and the order--disorder transition: at the
melting one has $\rho \sim a^{-2}$, while at the order--disorder
transition $\rho \sim R_a^{-2}\ll a^{-2}$. The position of the
upper critical point on the melting line is determined by the
condition that the density of dislocations in the disordered vortex
solid phase reaches a value typical for the liquid
(i.e., the dislocation spacing $R_a$ calculated on the melting
curve reduces to the vortex spacing $a$). It is this scenario that
is assumed in our paper.

  In the second scenario the upper critical point coincides with the
intersection point. In this case the above-mentioned point is
ordinary rather than singular for the free energy of the vortex
system, since a continuous phase transition separating the
disordered vortex solid phase from the vortex liquid does not seem
to exist.\cite{49,50} Thus, there is only {\it one} phase
transition line which describes both the melting and the
order--disorder transitions. This unified line originates from the
melting curve of clean superconductors, gradually evolving from it
as the strength of the quenched disorder increases. The ``upper
critical point'' is then simply the point where on the melting line
$R_a$ reduces to $a$ and the line experiences a bend. However, for
this scenario to occur, the calculated intersection point must lie in
the single vortex pinning region.
Otherwise, one obtains $R_a> R_c\gg a$ at this point and returns
to the first scenario.

  Our results argue in favor of
the first scenario since {\it in the framework of the used
approximations} the intersection points never reach the upper
region of single vortex pinning (see, e.g., Fig.~5). Note that
this scenario is also supported by recent numerical
simulations.\cite{51}

\section{Conclusions}

   In this paper, using the Lindemann criterion and
results of the collective pinning theory,\cite{27} we have derived
equations (19)-(25) which enable one to calculate the
order--disorder transition line $H_{dis}(t)$ with account of both
pinning-caused and thermal fluctuations in the whole temperature
interval of its existence. The boundaries of the single vortex
pinning region, $H_{sv}(t)$, $H_{sv}^{up}(t)$, and the melting
line $H_m(t)$ are also found from these equations. The equations
turn out to depend only on the Ginzburg number $Gi$ which
characterizes thermal fluctuations, on the strength of the
quenched disorder, $D$,
\[
  D={\epsilon \xi(0)\over L_c(0)}=
  \left(j_c(0) \over j_0(0) \right)^{1/2},
\]
and on the function $g_0(t)$ defined by the pinning mechanism and
by the temperature dependences of $\xi$ and $\lambda$, Appendix A.
For example, the pinning mechanism considered in
Ref.~\onlinecite{52} leads to a $g_0(t)$ described by Eq.~(A13),
while the $\delta T_c$ pinning\cite{27} results in Eq.~(A14).
Using our equations one can analyze the phase diagrams
of various superconductors. Moreover, our analysis in principle
allows to obtain information on the true form of the function
$g_0(t)$, i.e., on the mechanism of pinning in a superconductor.

  At small $Gi$ we obtain phase diagrams typical for low-$T_c$
superconductors. In this case the obtained results practically
coincide with the results of the simplified approach presented
in Sec.~III. We also analyze phase diagrams with $Gi\sim 10^{-2}$
since such values of $Gi$ are characteristic for the
${\rm YBa_2Cu_3O_{7-\delta}}$ superconductors. We consider both
a typical case where $g_0(t)$ decreases with $t$ [Eq.~(A13)]
and a case where $g_0(t)$ increases with $t$ [Eq.~(A14)].
In both these cases the obtained results qualitatively describe
the phase diagram of ${\rm YBaCuO}$ crystals and its evolution with
varying $D$, which can be caused by irradiation or by reduction of
the oxygen content in these crystals. However, we find that the
model of $\delta T_c$ pinning [Eq.~(A14)] more correctly reproduces
all the features of the experimental data for
${\rm YBa_2Cu_3O_{7-\delta}}$ crystals.

  Our results support the idea\cite{22} that the upper critical
point on the melting curve of high-$T_c$ superconductors does not
generally coincide with the intersection point of the melting and
the order--disorder lines (although the distance between them is
possibly small). We predict the position of this intersection
point in the $H-T$ plane. We suppose that the
temperature corresponding to this point is determined mainly by
the parameter $\nu$, Eq.~(29), i.e., by a combination of
the parameters $D$ and $Gi$.

 \acknowledgments

  G.P.M.~acknowledges the hospitality of the Max-Planck-Institut
f\"ur Metallforschung, Stuttgart. E.H.B. wishes to acknowledge
the hospitality of the Institute for Superconducting and
Electronic Materials, University of Wollongong, Australia,
where part of this work was performed, and financial support from
the Australian Research Council, IREX Program.

 \appendix{ }
\section{Estimates of the parameters}

 The results of this paper are based on the collective pinning
theory.\cite{27,32} It is assumed in this theory that the disorder
in the flux-line lattice is generated by point defects of size not
exceeding the coherence length $\xi(0)$. Here, in the framework of
three widely used models of pinning, we express the quantities
$\epsilon \xi /L_c$ and $\tilde T_{dp}^s$ through the
characteristics of the point defects, viz., their concentration
$n$ and mean radius $r_0$. These expressions enable us to get some
idea of the temperature dependences of these quantities and to
understand the changes of the phase diagram when $n$ and $r_0$ are
varied.

 In the first model,\cite{A1} pinning is due to the gain in the
condensation energy when a vortex core is located at a defect. This
gain $\epsilon_{pin}$ is of the order $(H_c^2/8\pi)4\pi r_0^3/3$,
where $H_c$ is the thermodynamic magnetic field and $r_0$ the radius
of the defect. Then one obtains the following estimates for the mean
elementary pinning force exerted by one point defect
 $f_{pin}\sim\epsilon_{pin}/\xi$:
\begin{equation} 
 f_{pin}\approx  \varepsilon_0\left({r_0 \over \xi}\right)^3,
\end{equation}
and for the single vortex collective pinning length\cite{27}
 $L_c=n^{-1/3}(\epsilon^2\varepsilon_0/f_{pin})^{\!2/3}$
[with $\epsilon=\lambda_{ab}/\lambda_c$,
 $\varepsilon_0 = (\Phi_0/4\pi\lambda_{ab})^2$]:
\begin{equation} 
   L_c\approx {\epsilon^{4/3} \over n^{1/3}}\left({\xi
   \over r_0}\right)^{\!2}.
\end{equation}
Using these estimates, we find the key parameter (7),
\begin{equation} 
 {\epsilon\xi \over L_c}\approx
 \left ({ nr_0^3 \over \epsilon}\right)^{\!1/3}{r_0 \over \xi},
\end{equation}
and the characteristic pinning energy\cite{27}
$\tilde T_{dp}^s$ described by the expression
 $\xi(f_{pin}^2nL_c\xi ^2)^{1/2}$,
\begin{equation} 
 \tilde T_{dp}^s\approx \epsilon^{2/3} \varepsilon_0(nr_0^3)^{1/3}r_0.
\end{equation}

 A different mechanism of pinning is due to the scattering of
quasiparticles by the defect as calculated in Ref.~\onlinecite{52}. A
scattering center facilitates deformations of the order parameter up
to distances of the order of the zero-temperature coherence length
$\xi(0)$. Hence, it is energetically advantageous for a vortex core
to sit at a scattering center. This mechanism leads to the pinning
force
 $f_{pin}\sim (H_c^2/8\pi)r_0^2\xi(0)/\xi$, and we obtain
 \begin{eqnarray} 
 f_{pin}&\approx & \varepsilon_0\left({r_0 \over
 \xi}\right)^{\!2} {\xi(0)\over \xi} \,, \\
 L_c&\approx &{\epsilon^{4/3} \over n^{1/3}}
 \left({\xi \over r_0}\right)^{\!4/3}
 \left ({\xi \over \xi(0)}\right )^{\!2/3}, \\
 {\epsilon\xi \over L_c}&\approx &
 \left ({ nr_0^3 \over  \epsilon}\right )^{\!1/3}
 \left ({r_0\xi^2(0) \over \xi^3}\right )^{\!1/3}, \\
 \tilde T_{dp}^s&\approx &\epsilon^{2/3}
 \varepsilon_0(nr_0^3)^{1/3}(r_0\xi^2(0))^{1/3}.
 \end{eqnarray}
Strictly speaking, formulas (A5)-(A8) are valid if
$\pi r_0^2\xi(0)n<1$. This type of pinning is sometimes
called $\delta l$ pinning.\cite{27} Note that equations (A5)-(A8)
and (A1)-(A4) have  the same  dependences  on $n$, $\epsilon$ and
temperature. They thus describe {\it two different contributions}
to one pinning mechanism, with the contribution of Eqs.~(A5)-(A8)
always dominating when  $r_0<\xi(0)$.

 If the pinning centers in ${\rm YBa_2Cu_3O_{7-\delta}}$ are
clusters of oxygen vacancies, it is useful to keep in mind that
\begin{equation} 
  {4\pi \over 3}r_0^3n=c\delta,
\end{equation}
where the constant $c$ equals $2$ if these clusters are formed by
${\rm YBa_2Cu_3O_{6.5}}$ phase. Thus, when $r_0$ decreases at a
fixed $\delta$, the parameters $\epsilon \xi/L_c$ and $T_{dp}^s$
also decrease.

 In high-$T_c$ superconductors the pinning can be due to spatial
fluctuations in the density of the oxygen vacancies, which
results in variations of $T_c$ over the sample. One has the
following estimates for this $\delta T_c$ pinning:\cite{27}
 \begin{eqnarray} 
 {\epsilon\xi \over L_c}&\propto &
 \left ({ n\xi \over \epsilon}\right)^{\!1/3}, \\
 \tilde T_{dp}^s&\propto &\epsilon^{2/3}
 \varepsilon_0(n\xi^4)^{1/3} ,
 \end{eqnarray}
where $n$ is the density of oxygen vacancies. This pinning
can occur when $n\epsilon \xi(0)^3 >1$.

 The above equations enable us to
estimate the temperature dependence of the quantities
$\epsilon \xi /L_c$ and $\tilde T_{dp}^s$. For this purpose, we
insert the expressions
\begin{equation} 
   {\xi(T)\over \xi(0)}={\lambda(T)\over \lambda(0)}=
   \left (1-{T^2\over T_c^2}\right)^{\!-1/2}
\end{equation}
in the appropriate formulas. In particular, it follows from
Eqs.~(10), (A3), (A7) that for $\delta l$ pinning
\begin{equation} 
   g_0(t)\equiv {1\over D}\left({\epsilon \xi(T)\over
   L_c(T)}\right) = (1-t^2)^{1/2} ,
\end{equation}
where $t\equiv T/T_c$, $D\equiv \epsilon \xi(0)/L_c(0)$, while
according to Eq.~(A10), one finds for $\delta T_c$ pinning
\begin{equation} 
   g_0(t)= (1-t^2)^{-1/6} .
\end{equation}


\newpage

 \begin{figure}[F1]
\epsfxsize= .98\hsize  \vskip 1.0\baselineskip \centerline{
\epsffile{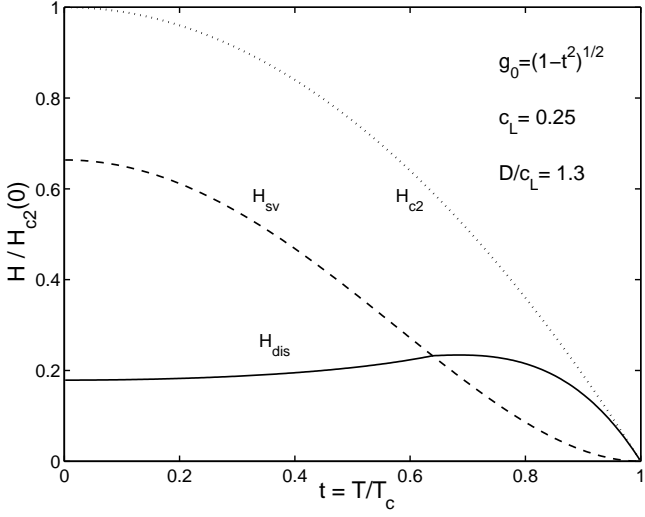}}
 \vspace{.1cm}
\caption{ The order--disorder line $H_{dis}(t)$ (solid line)
calculated from Eqs.~(4), (9), (A13) ($\delta l$ pinning)
for $c_L=0.25$ and $D/c_L =1.3$.
Simplified approach, without account of thermal fluctuations.
The boundary of the single vortex pinning regime, $H_{sv}(t)$, is
given by the dashed line, and the dotted line shows the mean-field
upper critical field $H_{c2}(t) =H_{c2}(0)(1-t^2)$.
 } \end{figure}

 \begin{figure}[F2]
\epsfxsize= .98\hsize  \vskip 1.0\baselineskip \centerline{
\epsffile{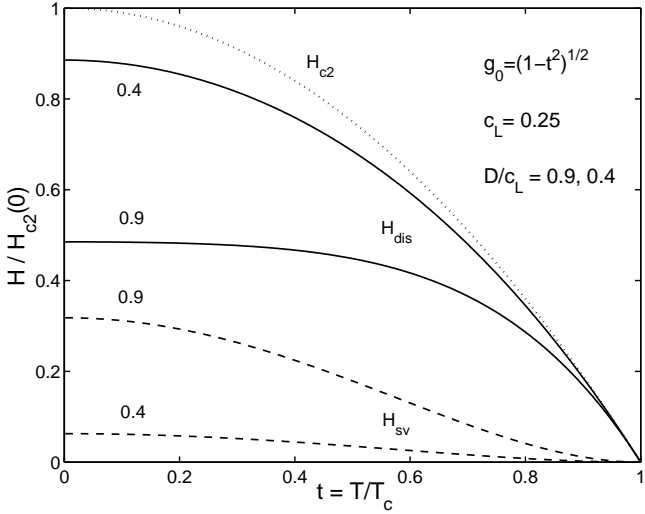}}
 \vspace{.1cm}
\caption{ As Fig.~1 but for $D/c_L = 0.9$ and 0.4. Note that for
$D/c_L < 1$ the order--disorder line $H_{dis}(t)$ does not
intersect the single vortex pinning boundary $H_{sv}(t)$.
 } \end{figure}

 \begin{figure}[F3]
\epsfxsize= .98\hsize  \vskip 1.0\baselineskip \centerline{
\epsffile{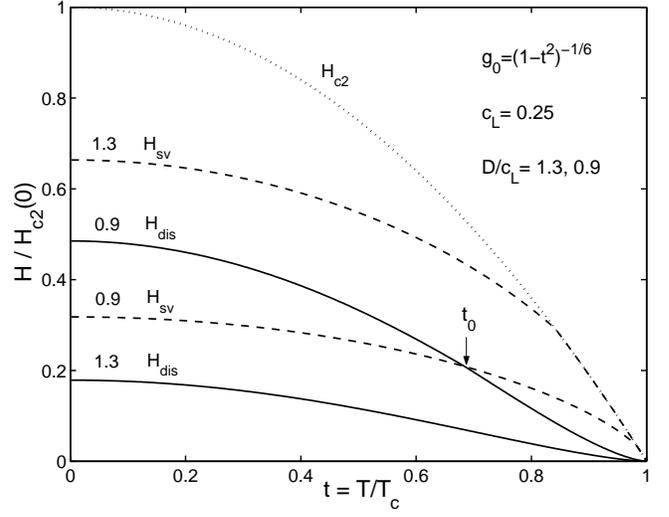}}
 \vspace{.1cm}
\caption{ The lines $H_{dis}(t)$ and $H_{sv}(t)$ as in Figs.~1, 2
but for $\delta T_c$ pinning, Eq. (A14), for
$D/c_L = 1.3$ and 0.9. Now $H_{dis}(t)$ and $H_{sv}(t)$ cross
at $t=t_0$ for $D/c_L <1$.
 } \end{figure}

 \begin{figure}[F4]
\epsfxsize= .98\hsize  \vskip 1.0\baselineskip \centerline{
\epsffile{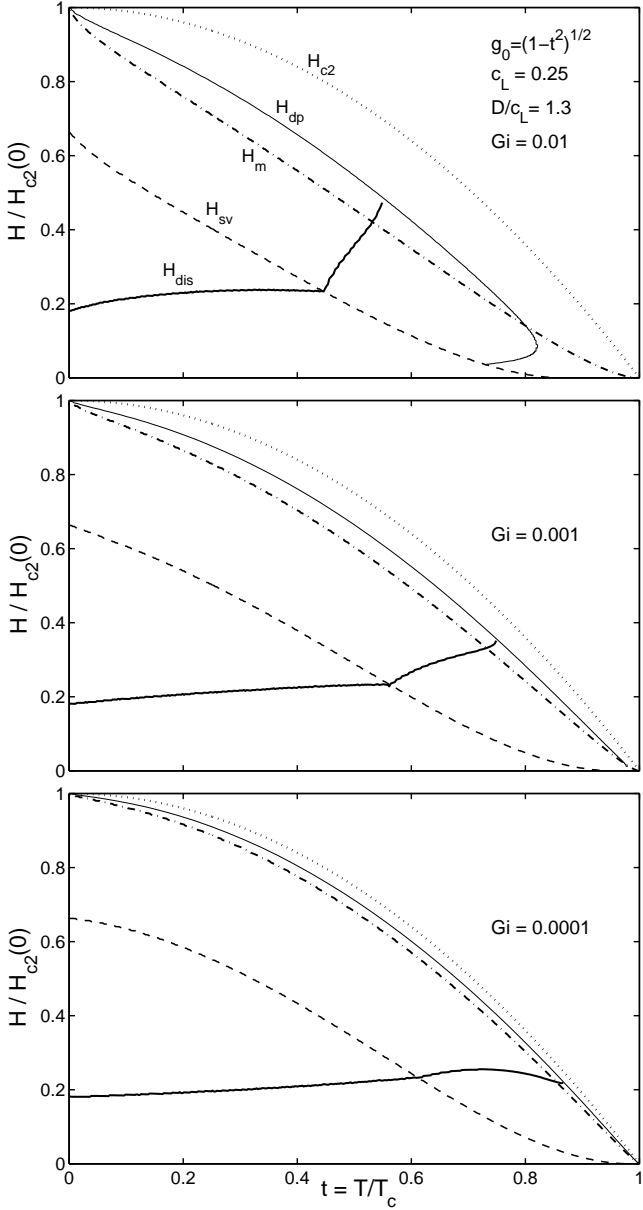}}
 \vspace{.1cm}
\caption{ The order--disorder line $H_{dis}(t)$ (bold solid line)
calculated from  Eqs.~(22), (24), (A13) ($\delta l$ pinning),
which account for thermal
fluctuations, for $c_L=0.25$, $D/c_L=1.3$, and for three values of
the Ginzburg number $Gi=0.01$, 0.001, and 0.0001 (from top to
bottom). Also shown are the boundary of the single vortex pinning
regime $H_{sv}(t)$, Eq.~(19) (dashed line), the mean-field
$H_{c2}(t) =H_{c2}(0)(1-t^2)$ (dotted line), the vortex-lattice
melting line $H_m(t)$, Eq.~(25) (dashed-dotted line), and the
depinning line $H_{dp}(t)$, Eqs.~(16)-(18) (thin solid line),
along which one formally has $u_T^2 =\xi^2$
and where $H_{dis}(t)$ ends. The short part of $H_{dis}$ above
$H_m$ has no physical meaning.
 } \end{figure}

 \begin{figure}[F5]
\epsfxsize= .98\hsize  \vskip 1.0\baselineskip \centerline{
\epsffile{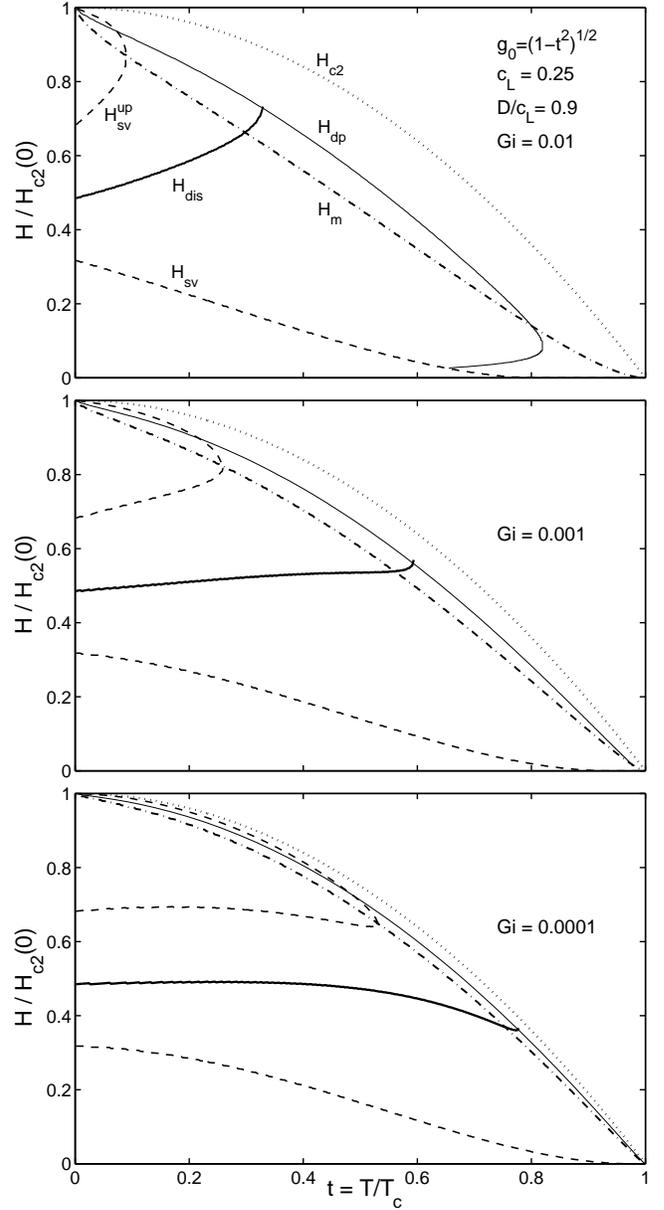}}
 \vspace{.1cm}
\caption{ As Fig.~4 but for different $D/c_L=0.9$. The dashed
lines show the boundaries of both the lower ($H_{sv}$)
and upper ($H_{sv}^{up}$) regions of single vortex pinning,
see Eq.~(21) and text.
 } \end{figure}

 \begin{figure}[F6]
\epsfxsize= .98\hsize  \vskip 1.0\baselineskip \centerline{
\epsffile{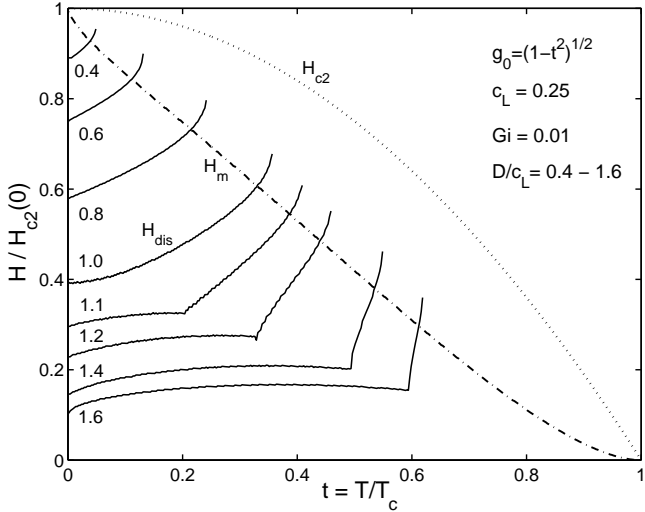}}
 \vspace{.1cm}
\caption{ The order--disorder lines $H_{dis}(t)$ (bold solid
lines) according to Eqs.~(22), (24), (A13) ($\delta l$ pinning)
for $c_L=0.25$, $Gi=0.01$ and several values $D /c_L= 0.4$ to 1.6.
The dash-dotted line is the vortex-lattice melting line $H_m(t)$,
Eq.~(25), and the dotted line indicates $H_{c2}(t)=H_{c2}(0)(1-t^2)$.
 } \end{figure}

 \begin{figure}[F7]
\epsfxsize= .98\hsize  \vskip 1.0\baselineskip \centerline{
\epsffile{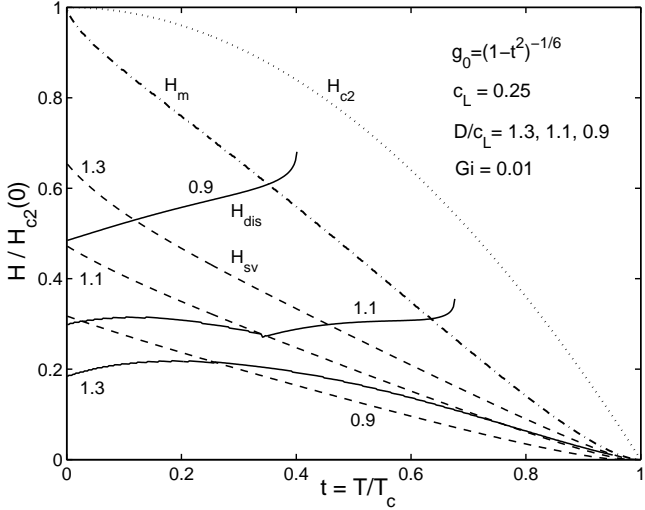}}
 \vspace{.1cm}
\caption{ The order--disorder lines $H_{dis}(t)$ (solid lines) as
in Fig.~6 but for $\delta T_c$ pinning, (A14), for $D /c_L=
1.3$, 1.1, and 0.9. The dashed lines are the single-vortex pinning
boundaries $H_{sv}(t)$, Eq.~(19), the dash-dotted line is the
melting line $H_m(t)$, Eq.~(25), and the dotted line indicates
$H_{c2}(t) =H_{c2}(0)(1-t^2)$.
 } \end{figure}

 \begin{figure}[F8]
\epsfxsize= .98\hsize  \vskip 1.0\baselineskip \centerline{
\epsffile{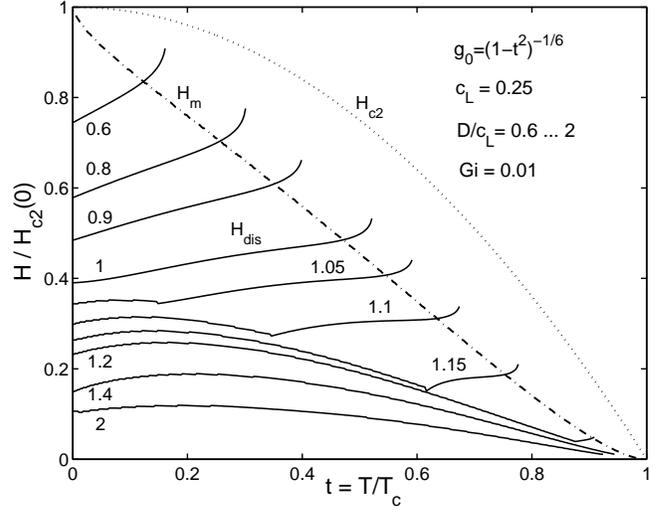}}
 \vspace{.1cm}
\caption{ The order--disorder lines $H_{dis}(t)$ (solid lines)
as in Fig.~(7) with (A14), but for many values $D /c_L= 0.6$ to 2.
 } \end{figure}

 \begin{figure}[F9]
\epsfxsize= .98\hsize  \vskip 1.0\baselineskip \centerline{
\epsffile{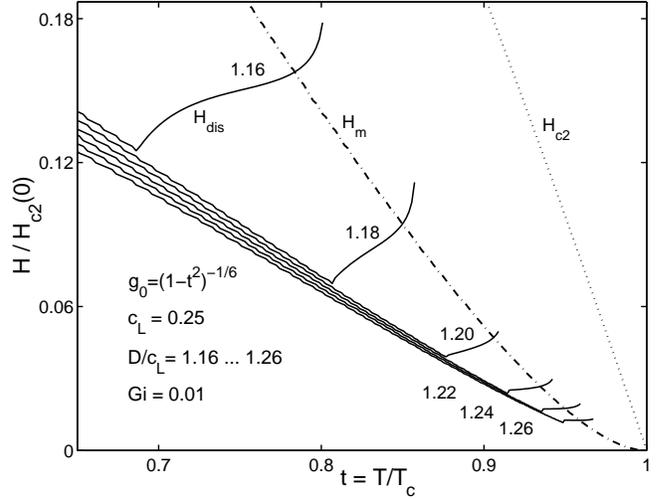}}
 \vspace{.1cm}
\caption{ As Fig.~8 but enlarged scale, for $D/c_L =1.16$ to
1.26. This figure simulates the evolution of the $H-T$ phase
diagram for ${\rm YBa_2Cu_3O_{7-\delta}}$ crystals when the
oxygen deficiency $\delta $ is changed.
 } \end{figure}

 \begin{figure}[F10]
\epsfxsize= .98\hsize  \vskip 1.0\baselineskip \centerline{
\epsffile{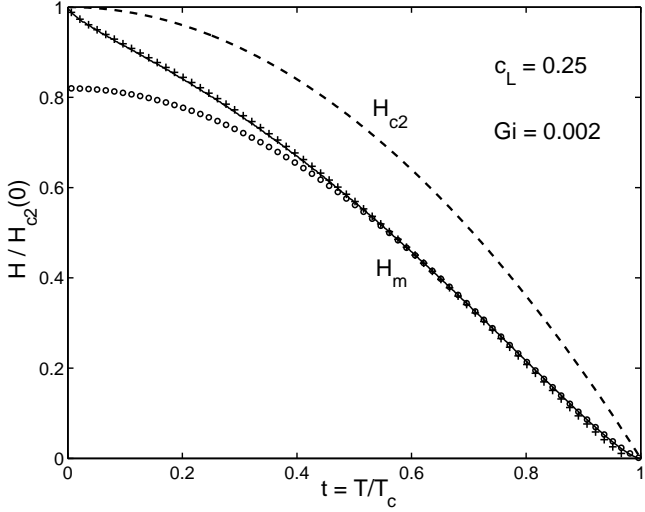}}
 \vspace{.1cm}
\caption{ The vortex-lattice melting-line $H_m(t)$ from Eq.~(25)
(solid line) and its approximations Eq.~(26) (crosses)  and
Eq.~(28) (circles) with $A=0.82$ and $\gamma=1.31$. The dashed
line indicates $H_{c2}(t) \propto 1-t^2$. Here $c_L=0.25$ and
$Gi=0.002$.
 } \end{figure}

 \begin{figure}[F11]
\epsfxsize= .98\hsize  \vskip 1.0\baselineskip \centerline{
\epsffile{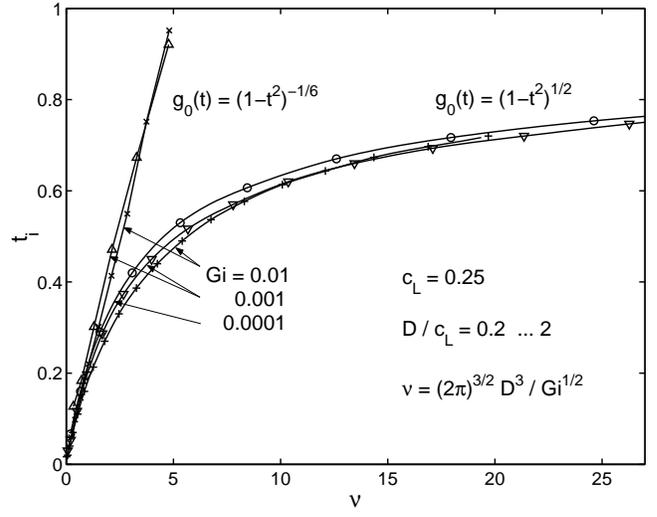}}
 \vspace{.1cm}
\caption{ The temperature $t_i$ where the melting line and the
order--disorder line cross, plotted versus the parameter
$\nu=(2\pi)^{3/2} D^3/Gi^{1/2}$. This temperature is found by
calculating the phase diagrams from Eqs.~(19)-(25) with $c_L=0.25$
and Eq.~(A13) (right 3 curves) or Eq.~(A14) (left 2 curves) for
$Gi=0.01$ (crosses), $Gi=0.001$ (triangles), and $Gi=0.0001$
(circles). The interval of $D/c_L$ actually used for each curve
can be calculated from $Gi$ and the $\nu$ interval.
 } \end{figure}

\end{multicols}
\end{document}